\documentclass[12pt]{article}
\def\be{\begin{equation}}
\def\ee{\end{equation}}
\def\ba{\begin{eqnarray}}
\def\ea{\end{eqnarray}}

\def\>{\rangle}
\def\<{\langle}

\def\n{\nonumber}

\def\<{\langle}
\def\>{\rangle}

\begin{document}
\begin{center}
{\Large\bf Quantum network architecture of tight-binding models with 
substitution sequences}\\
\medskip
Ilki~Kim and G\"{u}nter~Mahler\\
Institut f\"ur Theoretische Physik~I, Universit\"{a}t Stuttgart\\
Pfaffenwaldring 57, 70550 Stuttgart, Germany\\
phone: ++49-(0)711 685-5100, FAX: ++49-(0)711 685-4909\\
e-mail: ikim@theo.physik.uni-stuttgart.de
\end{center}

\begin{abstract}
 We study a two-spin quantum Turing architecture, in which discrete local
 rotations $\{\alpha_m\}$ of the Turing head spin alternate with quantum
 controlled NOT-operations. Substitution sequences are known to underlie 
 aperiodic structures. We show that parameter inputs $\{\alpha_m\}$ 
 described by such sequences can lead here to a quantum dynamics, 
 intermediate between the regular and the chaotic variant.
 Exponential parameter sensitivity characterizing chaotic quantum Turing 
 machines turns out to be an adequate criterion for induced quantum chaos 
 in a quantum network.
\end{abstract}

\section{Introduction}

Models described by one-dimensional Schr\"{o}dinger equations with 
quasi-periodic potentials display 
interesting spectra: this kind of potential is intermediate 
between periodic ones, leading to energy bands and 
extended states, and truly random potentials, which cause localisation 
\cite{OST83}. A super-lattice e.g. may be made of two species of 
doped semiconductors producing a one-dimensional chain of quantum wells. 
A qualitative model to describe the corresponding wave-functions 
is given by the tight-binding model, 
$\hat{H} \psi(n) = \psi(n+1) + \psi(n-1) + \lambda V(n) \psi(n)\,,\, 
\psi \in l^2(\mathbf{Z})$\,, where $V(n)$ represents the effect of 
quantum well~$n$\,, 
and $\lambda$ is a positive parameter playing the role 
of a coupling constant \cite{BOV95}. 
An interesting description results when this super-lattice is constructed 
by means of a deterministic rule. The simplest rule obtains when 
the two species alternate in a periodic way. 
But in general the rule will be aperiodic. One widely studied example is the 
Fibonacci sequence, which is quasi-periodic \cite{BOV95}: 
given two `letters' $a$ and $b$, one substitutes $a \to \xi(a) = ab$ 
and $b \to \xi(b) = a$. Iterating this rule on $a$ one thus generates the 
sequence $abaababaabaababa \cdots$\,, in which the frequency 
of $a$'s is given by the golden mean $(5^{1/2}-1)/2$. Other examples of such 
substitution rules are the Thue-Morse sequence (non quasi-periodic), 
which is obtained through the 
substitution $a \to \xi(a) = ab\,,\, b \to \xi(b) = ba$\, 
giving $abbabaabbaababba \cdots$\,, and the period-doubling sequence 
(non quasi-periodic), $a \to \xi(a) = ab\,,\, b \to \xi(b) = aa$ \cite{BOV95}. 
Accordingly, a piece-wise constant potential 
$\{ V_n; n \in \mathbf{Z} \}$ based on such such a rule, e.g. 
$V_1 = a, V_2 = b, V_3 = a,\, \cdots $\,, is called 
a `substitution potential'. The potential of the Fibonacci sequence has 
one-dimensional quasi-crystalline properties.

In recent years problems of quantum computing~(QC) and information processing 
have received increasing attention. To solve certain classes of problems 
in a potentially very powerful way, one tries to utilize in~QC the 
quantum-mechanical superposition principle and the (non-classical) 
entanglement. In the models of QC based on 
quantum Turing machines~(QTM) \cite{BEN82,DEU85}, 
the computation is characterized by sequences of unitary transformations 
(i.e. by the corresponding Hamiltonians~$\hat{H}$ acting during finite time 
interval steps). Benioff has studied the tight-binding model 
in a generalized QTM \cite{BEN97}, 
where QC is associated with different potentials at different steps as an 
`environmental effect'. These kinds of influences may 
introduce {\em{deterministic disorder}}\,
which would degrade performance by causing reflections 
at various steps and decay of the transmitted component \cite{LAN96}. 

Here we investigate an iterative map on qubits 
which can be interpreted as a QTM architecture \cite{KIM99}: 
local transformations of the Turing head controlled 
by a sequence of rotation angles $\{\alpha_m\}\,,\, m = 1, 2, \cdots$ 
(parameter inputs) alternate 
with a quantum-controlled NOT-operation~(QCNOT) with a second spin 
on the Turing tape. Those angles at steps $n = 2m-1$ are reminiscent of the 
potentials $V_m$ introduced before. 
In the present paper we will investigate the Fibonacci rule and the 
Thue-Morse case mainly with respect to the local dynamics of 
the Turing head, which will be shown to reflect 
the degree of `randomness' of the substitution sequences. 
The various types of aperiodic structures have been characterized 
up to now by the nature of their Fourier spectra only \cite{BOV95}. 

\section{Quantum Turing machine driven 
by substitution sequences}

The quantum network \cite{MAH98} to be considered here is composed 
of two pseudo-spins 
$|p\rangle\!^{(\mu)};\,p=-1,1;\,\mu=S,1$ 
(Turing-head $S$, Turing-tape spin $1$, see figure~\ref{QTM_chaos}) 
so that its network state $|\psi\rangle$ lives in the {four-dimensional} 
Hilbert space spanned by the product 
wave-functions $|j^{(S)} k^{(1)}\rangle = |jk\rangle$. 
Correspondingly, 
any (unitary) network operator can be expanded as a sum of product operators. 
The latter may be based on the $SU(2)$-generators, the 
Pauli matrices $\hat{\sigma}_j^{(\mu)},\, j=1,2,3$, together 
with the unit operator $\hat{1}^{(\mu)}$.

The initial state $|\psi_{0}\rangle$ will be taken to be 
a product of the Turing-head and tape wave-functions. 
For the discretized dynamical description of the QTM 
we identify the unitary operators $\hat{U}_{n}\,,\,n=1,2,3,\cdots$ 
(step number) with the local unitary transformation on the Turing head $S$, 
$\hat{U}_{\alpha_m}^{(S)}$, and the QCNOT on ($S,1$), $\hat{U}^{(S,1)}$, 
respectively, as follows:
\begin{eqnarray}
&&\hat{U}_{2m-1}\; =\; 
\exp{\left(-i \hat{\sigma}_1^{(S)} {\alpha_{m}}/2\right)}\label{us}\\
&&\hat{U}_{2m}\; =\; \hat{U}^{(S,1)}\; =\; 
\hat P_{-1,-1}^{(S)}\, \hat \sigma_1^{(1)} + 
\hat P_{1,1}^{(S)}\, \hat{1}^{(1)}\; =\; 
\left(\hat{U}^{(S,1)}\right)^{\dagger}\,,\label{ub}
\end{eqnarray}
where $P_{j,j}^{(S)} = |j\rangle\!^{(S)} {}^{(S)}\hspace*{-0.8mm}
\langle j|$ is a (local) projection operator, and 
the Turing head is externally driven by substitution sequences 
$\{\alpha_{m}\}$ specified by $\alpha_1, \alpha_2$\,. Here we restrict 
ourselves to the quasi-periodic 
Fibonacci - (qf) and Thue-Morse sequence (tm), respectively: 
\ba
&&\alpha_1^{\mbox{qf}} = \alpha_1\,,\, \alpha_2^{\mbox{qf}} = \alpha_2\,,\, 
\alpha_3^{\mbox{qf}} = \alpha_1\,,\, \alpha_4^{\mbox{qf}} = \alpha_1\,,\, 
\alpha_5^{\mbox{qf}} = \alpha_2\,,\, \cdots\label{fibo}\\
&&\alpha_1^{\mbox{tm}} = \alpha_1\,,\, \alpha_2^{\mbox{tm}} = \alpha_2\,,\, 
\alpha_3^{\mbox{tm}} = \alpha_2\,,\, \alpha_4^{\mbox{tm}} = \alpha_1\,,\, 
\alpha_5^{\mbox{tm}} = \alpha_2\,,\, \cdots\,.\label{tm}
\ea

First, we consider the reduced state-space dynamics of 
the head~$S$ and tape-spin~$1$, respectively, 
\begin{eqnarray}
\sigma_{j}^{(S)}(n)\;\, = &\mbox{Tr} \left( \hat{\rho}_n^{(S)}\, 
\hat{\sigma}_j^{(S)} \right)& =\;\, 
\langle\psi_{n}|\hat{\sigma}_{j}^{(S)} \otimes \hat{1}^{(1)}
|\psi_{n}\rangle\,,\nonumber\\
\sigma_{k}^{(1)}(n)\;\, = &\mbox{Tr} \left( \hat{\rho}_n^{(1)}\, 
\hat{\sigma}_k^{(1)} \right)& =\;\, 
\langle\psi_{n}|\hat{1}^{(S)} \otimes \hat{\sigma}_{k}^{(1)}|\psi_{n}
\rangle\,,\label{bloch}
\end{eqnarray}
where $|\psi_n\rangle$ is the total network state at step $n$, and 
$\sigma_{j}^{(\mu)}(n)$ are the respective Bloch-vectors. 
Due to the entanglement between the head and tape, both will, in 
general, appear to be in a `mixed-state', which means that the length of the 
Bloch-vectors in (\ref{bloch}) is less than $1$. 
However, for specific initial states 
$|\psi_0\rangle$ the state of head and tape will remain pure: 
As $|\pm\rangle\!^{(1)} = 
\frac{1}{\sqrt{2}}\left(|\hspace*{-1.mm}-\hspace*{-1.mm}1
\rangle\!^{(1)} \pm |1\rangle\!^{(1)}\right)$ are the eigenstates of 
$\hat{\sigma}_{1}^{(1)}$ with $\hat{\sigma}_{1}^{(1)} 
|\pm\rangle\!^{(1)} = \pm |\pm\rangle\!^{(1)}$, 
the QCNOT-operation $\hat{U}^{(S,1)}$ of equation~(\ref{ub}) cannot create 
any entanglement, irrespective of the head state 
$|\varphi\rangle\!^{(S)}$, i.e.
\begin{eqnarray}
\label{entangle}
\hat{U}^{(S,1)}\,|\varphi\rangle\!^{(S)} \otimes\,|+\rangle\!^{(1)}\,&=&\,
|\varphi\rangle\!^{(S)} \otimes\,|+\rangle\!^{(1)}\nonumber\\
\hat{U}^{(S,1)}\,|\varphi\rangle\!^{(S)} \otimes\,|-\rangle\!^{(1)}\,&=&\,
\hat{\sigma}_{3}^{(S)} 
|\varphi\rangle\!^{(S)} \otimes\,|-\rangle\!^{(1)}\,.
\end{eqnarray}
As a consequence, the state~$|\psi_n\rangle$ remains a product state for any 
step~$n$\, and initial product state 
$|\psi_0^{\pm}\rangle = |\varphi_0\rangle\!^{(S)} \otimes 
|\pm\rangle\!^{(1)}$\, with 
$|\varphi_0\rangle\!^{(S)} = \exp{
\left(-i \hat{\sigma}_1^{(S)} {\varphi_{0}}/2\right)}\, 
|\hspace*{-1.mm}-1\rangle\!^{(S)}$\,, 
so that the Turing head performs a pure-state trajectory 
(`primitive', see \cite{KIM99}) on 
the Bloch-circle $\left(\sigma_{1}^{(S)}(n)=0\right)$
\ba
&|\psi_{n}^{\pm}\rangle = |\varphi_{n}^{\pm}\rangle\!^{(S)} \otimes 
|\pm\rangle\!^{(1)}\,,
\;\;\;\;\left(\sigma_{2}^{(S)}(n|\pm)\right)^{2} + 
\left(\sigma_{3}^{(S)}(n|\pm)\right)^{2} = 1\,.\n&
\ea

Here $\sigma_{j}^{(S)}(n|\pm)$ denotes the Bloch-vector of the 
Turing head $S$ conditioned by the initial state $|\psi_0^{\pm}\rangle$. 
From the Fibonacci sequence (\ref{fibo}) and the property (\ref{entangle}) 
it is found for $|\varphi_{n}^{+}\rangle\!^{(S)} \otimes\,|+\rangle\!^{(1)}, 
n = 2m$, and $\varphi_0^{\pm} = \alpha_0 = 0$ that
\begin{equation}
\label{plus_fibo}
\sigma_{2}^{(S)}(2m|+) = \sin {\mathcal{C}}_{2m}(+)\,,\;\;\;\;
\sigma_{3}^{(S)}(2m|+) = -\cos {\mathcal{C}}_{2m}(+)\,,
\end{equation}
and\, $\sigma_{k}^{(S)}(2m-1|+) = \sigma_{k}^{(S)}(2m|+)$\,, where 
the cumulative rotation angle is
\ba
&{\mathcal{C}}_{2m}(+)\; =\; 
{\displaystyle \sum_{j=1}^{m} \alpha_{j}^{\mbox{qf}}}\; =\; 
\alpha_1\, m\, +\, (\alpha_2 - \alpha_1)\, m'\,,&\n
\ea
with\, $m'\, (\leq m)$ being the total number of angles\, $\alpha_2$ up to 
step $2m$. For the cumulative rotation angle 
${\mathcal{C}}_{n}(-)$ up to step $n$ we utilize the following recursion 
relations
\ba
&{\mathcal{C}}_{2m}(-) = -{\mathcal{C}}_{2m-1}(-)\,,\;\;\;\;
{\mathcal{C}}_{2m-1}(-) = \alpha_{m}^{\mbox{qf}} + 
{\mathcal{C}}_{2m-2}(-)\,.&\n
\ea
Then it is easy to verify that for 
$|\varphi_{n}^{-}\rangle\!^{(S)} \otimes\,|-\rangle\!^{(1)}$ and 
$\varphi_0^{\pm} = \alpha_0 = 0$
\be
\label{cumulative}
\left| {\mathcal{C}}_{n}(-) \right|\; \leq \; 2\, 
\max \left( |\alpha_1|\, ,\, |\alpha_2| \right)\, =:\, M\,,
\ee
yielding $\sigma_{2}^{(S)}(n|-) = \sin {\mathcal{C}}_{n}(-),\,
\sigma_{3}^{(S)}(n|-) = -\cos {\mathcal{C}}_{n}(-)$.

From any initial state, 
$|\psi_{0}\rangle = a^{(+)}|\varphi_{0}^{+}\rangle\!^{(S)} \otimes 
|+\rangle\!^{(1)} + a^{(-)}|\varphi_{0}^{-}\rangle\!^{(S)} \otimes 
|-\rangle\!^{(1)}$, we then obtain at step $n$
\ba
&|\psi_n\rangle = a^{(+)}|\varphi_{n}^{+}\rangle\!^{(S)} \otimes\,
|+\rangle\!^{(1)}\,+\,a^{(-)}|\varphi_{n}^{-}\rangle\!^{(S)} \otimes\,
|-\rangle\!^{(1)}&\n
\ea
and, observing the orthogonality of the $|\pm\rangle\!^{(1)}$, 
\begin{equation}
\label{super}
\sigma_{k}^{(S)}(n) = |a^{(+)}|^{2}\,\sigma_{k}^{(S)} (n|+)\,+\,
|a^{(-)}|^{2}\,\sigma_{k}^{(S)} (n|-)\,.
\label{lambda_S}
\end{equation}
This trajectory of the Turing-head~$S$ represents a non-orthogonal pure-state 
decomposition. By using (\ref{plus_fibo}), (\ref{cumulative}), (\ref{super}) 
$\left( \mbox{with}\; a^{(+)} = a^{(-)} = 1/\sqrt{2} \right)$ 
we finally get for 
$|\psi_0\rangle = |-1\rangle\!^{(S)} \otimes\,|-1\rangle\!^{(1)}$
\begin{eqnarray}
\label{chaotic_driving}
\left(\sigma_{2}^{(S)}(n),\,\sigma_{3}^{(S)}(n)\right)\;=\;
\cos {\mathcal{B}}_n \cdot 
\left( \sin {\mathcal{A}}_n ,\,-\cos {\mathcal{A}}_n \right)\,,
\end{eqnarray}
where $({\mathcal{C}}_{n}(+) - M)/2 \leq 
{\mathcal{A}}_n = ({\mathcal{C}}_{2m}(+) + {\mathcal{C}}_{2m}(-))/2\,,\; 
{\mathcal{B}}_n = ({\mathcal{C}}_{2m}(+) - {\mathcal{C}}_{2m}(-))/2 \leq 
({\mathcal{C}}_{2m}(+) + M)/2$\,, $n = 2m$\, or\, $2m-1$. 
Thus the expression~(\ref{chaotic_driving}) indicates that for the local 
dynamics of the Turing head in the `non-classical' regime 
the cumulative control loss due to any small 
perturbation $\delta$ of the given 
$\alpha_1^{\mbox{qf}}, \alpha_1^{\mbox{qf}}$ grows 
at most linearly with $n$, so that all periodic orbits on the plane 
$\left\{0, \sigma_{2}^{(S)}, \sigma_{3}^{(S)}\right\}$ are stable 
(see figure~\ref{stability}$a$ - $c$), 
as in the case of the `regular' control $\alpha_m = 
\alpha_1$ \cite{KIM99}. This may be contrasted with the chaotic 
Fibonacci-rule (cf), $\alpha_{m+1}^{\mbox{cf}} = \alpha_m^{\mbox{cf}} + 
\alpha_{m-1}^{\mbox{cf}}$ 
(Lyapunov exponent: $\ln\, (1 + \sqrt{5})/2 > 0$)\,, which can be interpreted 
as a temporal random (chaotic) analogue to one-dimensional chaotic potentials 
\cite{IKI99}: each step $\alpha_m$ is controlled 
by the cumulative information of the two previous steps. 
For a small perturbation of the initial phase angle $\alpha_0$ 
the cumulative angles ${\mathcal{A}}_m, {\mathcal{B}}_m$\,, respectively, 
grow exponentially with $m$, and so do 
the deviation terms $\Delta {\mathcal{C}}_{2m}^{\mbox{cf}}(\pm) = 
{\mathcal{C}}_{2m}^{\mbox{cf}\,'}(\pm) - 
{\mathcal{C}}_{2m}^{\mbox{cf}_{\mbox{po}}}(\pm)$ 
from the periodic orbits (po). Thus the total cumulative control loss induced 
by the perturbation can show chaotic quantum behaviour on the Turing head 
\cite{IKI99}.

For the Thue-Morse control (\ref{tm}) we easily find that for 
$|\psi_0\rangle = |-1\rangle\!^{(S)} \otimes\,|-1\rangle\!^{(1)}$ and $n = 8m$ 
\ba
&{\mathcal{C}}_{n}(+)\; =\; 2\, (\alpha_1 + \alpha_2)\, m\,,\;\;\;\; 
{\mathcal{C}}_{n}(-)\; =\; 0\,,&\n
\ea
respectively, which is very similar to the result of the 
`regular' machine with 
$\alpha_m^{\mbox{reg}} = (\alpha_1 + \alpha_2)/2$, implying 
${\mathcal{C}}_{8m}^{\mbox{reg}}(+) = 2 (\alpha_1 + \alpha_2)$ and 
${\mathcal{C}}_{8m}^{\mbox{reg}}(-) = 0$. 
In all cases considered we thus find characteristic local invariants 
with respect to the Turing head (figure~\ref{stability}$d$).

\section{Parameter sensitivity}

The distance between density operators, $\hat{\rho}$ and $\hat{\rho}'$, 
defined by the so-called Bures metric \cite{HUE92}
\ba
&D_{\rho \rho'}^{2} := 
\mbox{Tr} \left\{(\hat{\rho} - \hat{\rho}')^2\right\}\,.&\n
\ea
lies, independent of the dimension of the Liouville space, between 
0 and 2. For pure states we can rewrite
\ba
&&D^2\; =\; 2\,(1 - |\langle\psi|\psi'\rangle|^2)\; =\; 2\,( 1 - O')\n\\
&&O' := |\langle \psi_0(\delta)| \hat{U}^{\dagger}(\delta)\, 
\hat{U}(0) |\psi_0(0) \rangle|^2\,,\n
\ea
where the perturbed unitary evolution, $\hat{U}(\delta)$, connects the 
initial state, $|\psi_0(\delta)\rangle$, and $|\psi'\rangle$. 
This metric can be applied likewise to the total-network-state space 
or any subspace. In any case it is a convenient additional means to 
characterize various QTMs: for the regular case, 
$\alpha_m^{\mbox{reg}} = \alpha_1$ (Lyapunov exponent $= 0$), 
and any initial perturbation $\delta$ for $\hat{\rho}'$ 
the distance remains almost constant \cite{IKI99}; 
for the chaotic Fibonacci rule (cf), on the other hand, $\left( \alpha_m 
(\hat{\rho}) = \alpha_m^{\mbox{cf}}\,,\, \alpha_m (\hat{\rho}') = 
\alpha_m (\hat{\rho}) + \delta_{m}^{\mbox{cf}} \right.$\,, 
where $\delta_{m}^{\mbox{cf}}$ is the cumulative perturbation of the angle\, 
$\alpha_m^{\mbox{cf}}$\, at step $\left. n = 2m - 1 \right)$\, 
we obtain an initial exponential sensitivity \cite{IKI99}. 
In the case of the present substitution sequences 
we observe for a small perturbation of the given $\alpha_1, \alpha_2$ 
no initial exponential sensitivity in the evolution of $D^2$, 
which confirms that any periodic orbit is stable (figure~\ref{evolve}$a$). 
Finally we display the evolution of $D^2$ for the total network 
state $|\psi_n\rangle$, which also shows no exponential sensitivity 
(figure~\ref{evolve}$b$\,, cf. figure~3$c$ in \cite{IKI99}). 
The respective distances for tape-spin $1$ are similar to those shown. 
The corresponding behaviour under the Thue-Morse control is qualitatively 
the same. This parameter-sensitivity \cite{PER91} has been proposed as a 
measure to distinguish quantum chaos from regular quantum dynamics. 
From the results of the present analysis and those in \cite{IKI99} 
satisfying this criterion we conclude that, indeed, only classical chaotic 
input makes the quantum dynamics in QTM architectures chaotic, too 
(figure~\ref{QTM_chaos}).

\section{Summary}

In conclusion, we have studied the quantum dynamics of a small 
QTM driven by substitution sequences based on a decoherence-free Hamiltonian. 
As quantum features 
we utilized the superposition principle and the physics of entanglement. 
Quantum dynamics manifests itself in the superposition and entanglement of 
a pair of `classical' (i.e. disentangled) state-sequences. 
The generalized QTM under this kind of control 
connects two fields of much recent interest, quantum computation 
and motion in one-dimensional structures with `deterministically aperiodic' 
potential distributions. No chaotic quantum dynamics results in this case, as 
shown by the lack of exponential parameter-sensitivity. 
Local invariants leading to one-dimensional point manifolds (patterns) exist 
for $\alpha_1 = \alpha_2$\, only. For $\alpha_1 \ne \alpha_2$ a continuous 
destruction of these patterns sets in (figure~\ref{stability}$b$\,; hardly 
visible yet in figure~\ref{stability}$a$). 
This reminds us of the disappearing of KAM tori in the classical phase space 
resulting from a small perturbation (see e.g. \cite{ALL97}). 
Patterns in reduced 
Bloch-planes $\left\{0, \sigma_{2}^{(\mu)}, \sigma_{3}^{(\mu)}\right\}$ 
(a quantum version of a Poincar\'{e}-cut) should thus be similarly useful to 
characterize quantum dynamics in a broad class of quantum networks. 
Due to the entanglement, we can see regular, chaotic, and intermediate 
quantum dynamics, respectively. 
Furthermore, the parameter sensitivity gives a sensitive 
criterion for testing induced quantum chaos in a pure quantum regime. 
This might be contrasted with the usual quantum chaology, 
which is concerned essentially only with semiclassical spectrum analysis of 
classically chaotic systems 
(e.g. level spacing, spectral rigidity) \cite{BER85}. 
It is expected that a QTM architecture with a larger number of 
pseudo-spins on the Turing tape would still exhibit the same type of 
dynamical behaviour under the corresponding driving conditions.

\section{Acknowledgements}

We thank J.~Gemmer, A.~Otte, P.~Pangritz, and F.~Tonner 
for fruitful discussions. 
One of us (I.~K.) is grateful to Prof. P.~L.~Knight, Dr. M.~B.~Plenio, and 
their co-workers at Imperial College for their hospitality and 
to the European Science Foundation (Quantum information theory and quantum 
computation) for financial support during his visit.

\begin{figure}
\refstepcounter{figure}\label{QTM_chaos}
\refstepcounter{figure}\label{stability}
\refstepcounter{figure}\label{evolve}
\vspace*{25.5cm}
\hspace*{-1.55cm}
\includegraphics{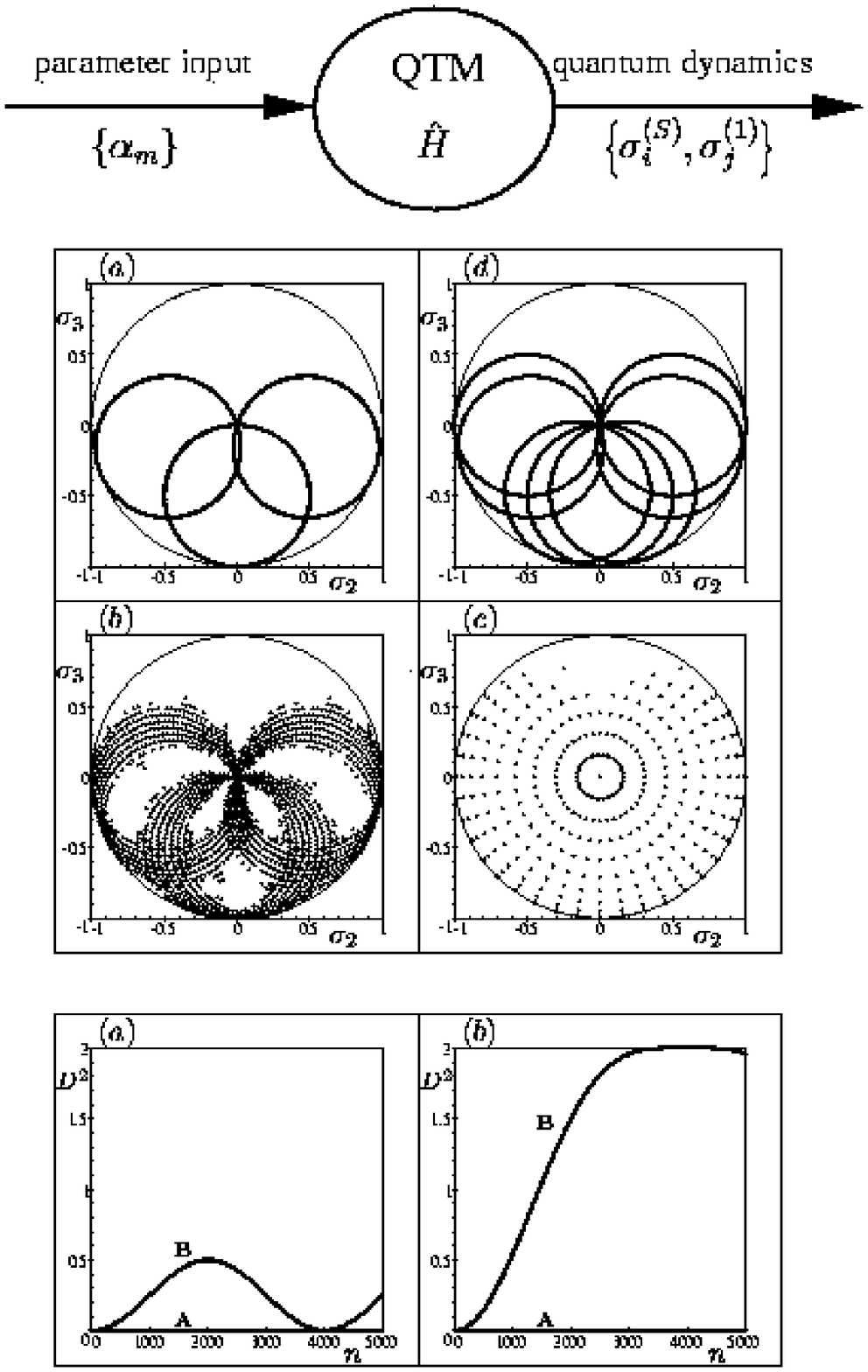}
\vspace*{-3cm}
\end{figure}

\newpage
Figure~\ref{QTM_chaos}: 
Input-output-scheme of our quantum Turing machine~(QTM).
\vspace*{0.5cm}

Figure~\ref{stability}: Turing-head patterns 
$\left\{0, \sigma_{2}(n), \sigma_{3}(n)\right\}$ for initial state 
$|\psi_{0}\rangle = 
|-1\rangle\!^{(S)} \otimes$\linebreak
$|-1\rangle\!^{(1)}$ under the control of substitution sequences: 
($a$) quasi-periodic Fibonacci (qf) with $\alpha_1 = 
\frac{2}{5} \pi\,,\, \alpha_2 = \alpha_1 + 0.0005 \pi$\,, 
($b$) as in $a$) but for $\alpha_2 = \alpha_1 + 0.03 \pi$\,, 
($c$) as in $a$) but for $\alpha_2 = \alpha_1 + 0.05 \pi$\,; 
($d$) Thue-Morse (tm) control with $\alpha_1 = \frac{2}{5} \pi\,,\, 
\alpha_2 = \alpha_1 + 0.1001 \pi$\,. 
For each simulation the total step number is $n=10000$.
\vspace*{0.5cm}

Figure~\ref{evolve}: Evolution of the (squared) distance $D_{\rho \rho'}^2$ 
between the perturbed, $\hat{\rho}'$\,, and the reference QTM 
state, $\hat{\rho}$\,, under the quasi-periodic Fibonacci (qf) control: 
($a$) for Turing head; ($b$) for total network state $|\psi_n\>$\,. 
For $\hat{\rho}$\, we take $|\psi_{0}\rangle = 
|-1\rangle\!^{(S)} \otimes\,|-1\rangle\!^{(1)}$ and $\alpha_1 = 
\frac{2}{5} \pi\,, \, \alpha_2 = \alpha_1 + 0.03 \pi$\,. 
Line {\bf A}: $|\psi_{0}'\rangle = \exp{
\left(-i \hat{\sigma}_1^{(S)} {\delta}/2\right)} |\psi_{0}\rangle$ 
for $\hat{\rho}'$\,, $\delta = 0.001$. 
Line {\bf B}: $|\psi_{0}'\rangle = |\psi_{0}\rangle$\,, but 
$\alpha_{1}' = \alpha_1 + 0.001 \pi\,,\, 
\alpha_{2}' = \alpha_2 + 0.001 \pi$\, for $\left(\hat{\rho}'\right)$\,.
\end{document}